\documentclass{aa}
\usepackage{psfig}
\usepackage{txfonts}
\usepackage{natbib}
\usepackage{amssymb}
\bibpunct{(}{)}{;}{a}{}{,}

\def\pcs{ph~cm$^{-2}$s$^{-1}$ }
\def\pcsk{ph~cm$^{-2}$s$^{-1}$keV$^{-1}$ }
\def\ecs{erg~cm$^{-2}$s$^{-1}$ }

\def\chir{$\chi^2_r$ }

\begin{document}
\bibliographystyle{aa}

\title{INTEGRAL survey of the Cassiopeia region in hard X rays}

\author{P.R. den Hartog\inst{1}
  \and W. Hermsen\inst{1,2}
  \and L. Kuiper\inst{1}
  \and J. Vink\inst{3,1}
  \and J.J.M. in 't Zand\inst{1,3}
  \and W. Collmar\inst{4}}

\offprints{P.R. den Hartog}
\mail{Hartog@sron.nl}
 
\institute{SRON Netherlands Institute for Space Research,
  Sorbonnelaan 2, 3584 CA Utrecht, The Netherlands
  \and Sterrenkundig Instituut Anton Pannekoek, University of Amsterdam,
  Kruislaan 403, 1098 SJ Amsterdam, The Netherlands
  \and Sterrekundig Instituut, Universiteit Utrecht, 
  P.O. Box 80000, 3508 TA Utrecht, The Netherlands 
  \and Max-Planck-Institut f\"ur extraterrestrische Physik, 
  PO Box 1603, 85740 Garching, Germany}

\date{Received 2005 December 16; accepted 2006 January 19}

\abstract{We report on the results of a deep 1.6 Ms INTEGRAL
observation of the Cassiopeia region performed from December 2003 to
February 2004. Eleven sources were detected with the imager IBIS-ISGRI
at energies above 20 keV, including three new hard X-ray sources. Most
remarkable is the discovery of hard X-ray emission from the anomalous
X-ray pulsar 4U~0142+61, which shows emission up to $\sim$150 keV with
a very hard power-law spectrum with photon index $\Gamma = 0.73 \pm
0.17$. We derived flux upper limits for energies between 0.75~MeV and
30~MeV using archival data from the Compton telescope COMPTEL. In
order to reconcile the very hard spectrum of 4U~0142+61 measured by
INTEGRAL with the COMPTEL upper limits, the spectrum has to bend or
break between $\sim$75 keV and $\sim$750 keV.  1E~2259+586,
another anomalous X-ray pulsar in this region, was not
detected. INTEGRAL and COMPTEL upper limits are provided. The new
INTEGRAL sources are IGR~J00370+6122 and
IGR~J00234+6144. IGR~J00370+6122 is a new supergiant X-ray binary with
an orbital period of $15.665 \pm 0.006$ days, derived from RXTE
All-Sky Monitor data. Archival BeppoSAX Wide-Field Camera data yielded
four more detections. IGR~J00234+6144 still requires a proper
identification. Other sources for which INTEGRAL results are presented
are high-mass X-ray binaries 2S~0114+650, $\gamma$~Cas,
RX~J0146.9+6121 and 4U~2206+54, intermediate polar V709~Cas and
1ES~0033+595, an AGN of the BL-Lac type. For each of these sources the
hard X-ray spectra are fitted with different models and compared with
earlier published results.

}

\maketitle

\section{Introduction}
\label{sec:intro}

The {\em INTErnational Gamma-Ray Astrophysics Laboratory} INTEGRAL
\citep{Winkler03_IGR} is ESA's currently operational space-based hard
X-ray/soft gamma-ray telescope. Since its launch in October 2002 it
has been observing the sky at photon energies between 3~keV and
2~MeV. Aboard INTEGRAL there are two main hard X-ray/soft gamma-ray
instruments, the imager IBIS \citep{Ubertini03_IBIS} and the
spectrometer SPI \citep{Vedrenne03_SPI}, supplemented by two X-ray
monitors, JEM-X \citep{Lund03_JMX} and an optical monitor, OMC
\citep{Mas-Hesse03_OMC}. These instruments are all wide-field
instruments with observational coverage over a broad spectral band. In
this work only INTEGRAL results from the low-energy detector of IBIS,
called ISGRI \citep{Lebrun03_ISGRI}, will be presented.  IBIS-ISGRI is
a coded-mask instrument with arcminute-resolution \citep[12\arcmin \,
FWHM;][]{Gros03_IBIS} imaging capabilities in the energy range from
$\sim$20~keV to $\sim$300~keV. Its field of view (FOV) is $29^\circ
\times 29^\circ$ (full width zero response). The most sensitive part
of the FOV of a coded-mask instrument is the fully-coded FOV, which is
$9^\circ \times 9^\circ$ for IBIS-ISGRI \citep[see][ for detailed
information on coded-mask instruments]{IntZand92_codedmask}.

The main targets of our INTEGRAL observations are the young Galactic
supernova remnants Cassiopeia~A and Tycho. In this paper these sources
will not be addressed in detail; we refer to \citet{Vink05_casa} and
\citet{Renaud04_tycho} for more detailed studies. This paper presents
results on compact hard X-ray sources detected in the deep 1.6-Ms
observation of the Cassiopeia region. In this direction of the
Galactic plane the line of sight crosses the Perseus arm at a distance
of $\sim$3 kpc. One may expect that the majority of the sources are
likely to be associated with this spiral arm. The detected sources
are: (Anomalous X-ray Pulsar; AXP) 4U~0142+61, (high-mass X-ray
binaries) 2S~0114+650, $\gamma$~Cas, RX~J0146.9+6121, 4U~2206+54, the
newly discovered IGR~J00370+6122, (intermediate polar) V709~Cas,
(BL-Lac) 1ES~0033+595, and the currently unidentified newly discovered
INTEGRAL source IGR~J00234+6141. INTEGRAL-flux upper limits are
provided for some transient sources and a second AXP (1E~2259+586) in
this field.

For the two AXP sources, 0.75--30 MeV archival data from COMPTEL
\citep{Schoenfelder93_comptel} are reanalyzed, delivering constraining
upper limits for 4U~0142+61. In an attempt to reveal the nature of
IGR~J00370+6122 archival data from the RXTE All-Sky Monitor and the BeppoSAX
Wide-Field Cameras have been studied, as well as a Target of
Opportunity (ToO) observation with RXTE PCA.

\section{INTEGRAL observations and analysis}
\label{sec:obs}
\begin{table}
\renewcommand{\tabcolsep}{1.7mm}
\caption[]{Summary of our INTEGRAL observations of the Cassiopeia
region. For every observation are given the revolution number, year,
begin and end time ($t_{begin}$ and $t_{end}$) and the exposure time
($t_{exp}$).}

\begin{tabular}{l l l@{ } l l l@{ }l l r@{.}l}

\vspace{-3mm}\\

\hline
\hline
Rev. & 
year &

\multicolumn{2}{l}{date} &
$t_{\mathrm{begin}}$ &
\multicolumn{2}{l}{date} &
$t_{\mathrm{end}}$ &
\multicolumn{2}{r}{$t_{\mathrm{exp}}\,(\rm{ks})$}\\

\hline

142 & 2003 & Dec. & 12&15:46:50 & Dec. & 14&21:11:53 & 178&2\\
143 & 2003 & Dec. & 15&07:19:13 & Dec. & 17&21:16:51 & 206&8\\
144 & 2003 & Dec. & 18&07:15:25 & Dec. & 20&15:00:42 & 187&0\\
145 & 2003 & Dec. & 21&16:36:36 & Dec. & 23&20:59:51 & 173&8\\
146 & 2003 & Dec. & 24&18:50:50 & Dec. & 26&20:23:19 & 165&0\\
147 & 2003 & Dec. & 27&06:58:07 & Dec. & 29&20:43:13 & 206&8\\
148 & 2003 & Dec. & 30&06:51:06 & Jan. & 01&15:10:42 & 189&2\\
161 & 2004 & Feb. & 07&18:13:47 & Feb. & 09&18:17:17 & 163&2\\
162 & 2004 & Feb. & 10&17:37:54 & Feb. & 12&18:06:20 & 163&2\\

\hline
\end{tabular}

\label{tab:obs}
\end{table}

\begin{figure*}
\psfig{figure=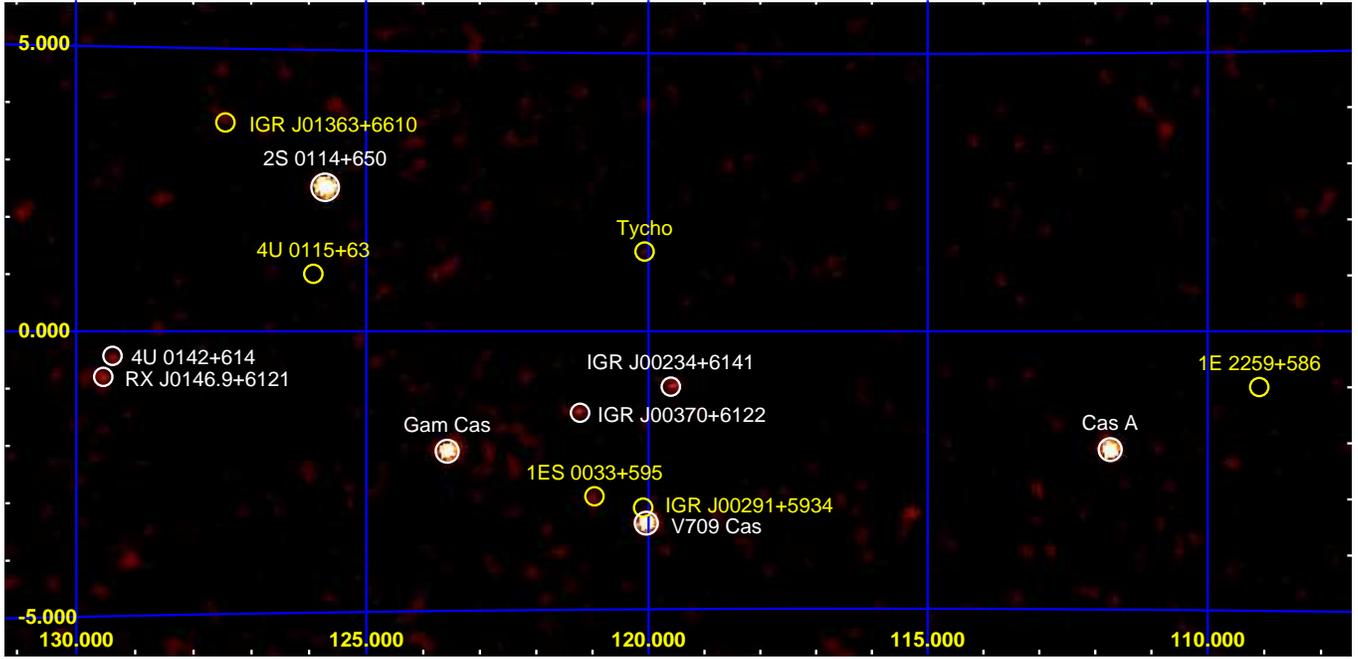,width=18cm,angle=-90,clip=t}
\caption{Cassiopeia-region significance map in Galactic coordinates
$(l,b)$ as seen with IBIS-ISGRI in the 20--50 keV energy range. All
excesses with positive significances are drawn. Indicated with white
circles and annotation are the sources significantly detected during
these observations. Indicated with yellow circles and annotation are
transient sources which have been weakly detected in this work and/or
detected with INTEGRAL at other epochs. AXP 1E~2259+586, not yet
detected, is a good candidate for future study (see section
\ref{sec:2259}). The weakest (marginally) detected source is Tycho
with $3.5\sigma$, while the brightest source in this field,
2S~0114+650, is detected with $43.1\sigma$
\label{fig:field20-50}}
\end{figure*}
\begin{figure*}
\psfig{figure=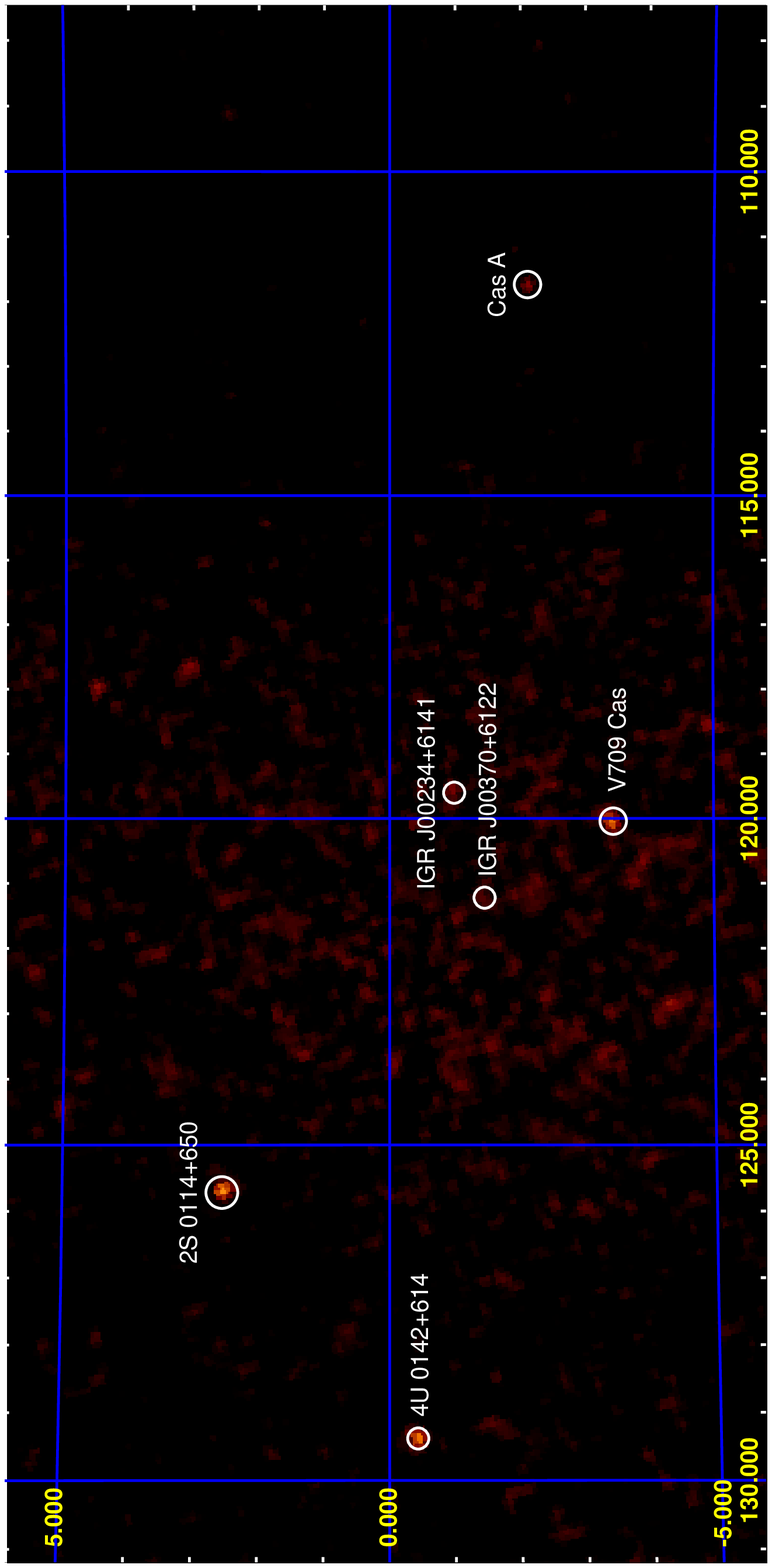,width=18cm,angle=-90,clip=t}
\caption{Cassiopeia region presented in the same format as in
Fig.~\ref{fig:field20-50} in the 50--100 keV energy range. All
detected sources are indicated with white circles and
annotation. Several sources are no longer apparent in this energy
range. IGR~J00370+6122 is not visible in this total-exposure map, but
is detected during two single revolutions (see
Sect.~\ref{sec:0037}). Some residual artefacts/systematics can be seen
in the center of this map.
\label{fig:field50-100}}
\end{figure*}

Our observations of the Cassiopeia region were made during nine
orbital revolutions between December 12, 2003 and February 12, 2004
(see Table~\ref{tab:obs} for details). INTEGRAL performed the
observations in a standard rectangular ($5 \times 5$) dithering mode
\citep{Jensen03_igr}.  The data provided by the INTEGRAL Science Data
Centre \citep[ISDC;][]{Courvoisier03_isdc} were screened for
Solar-flare events and erratic count-rate fluctuations due to passages
through the Earth's radiation belts. After screening, the net exposure
of the fully-coded FOV was 1.6 Ms, from 677 pointings called Science
Windows (ScWs). The data were reduced using the Off-line Scientific
Analysis (OSA) software package version 4.2 \citep[see][ for
IBIS-ISGRI scientific data analysis]{Goldwurm03_OSA}.  For each ScW
deconvolved IBIS-ISGRI sky maps were created in different energy
ranges. These deconvolved sky maps were combined into mosaic maps.

The mosaics were examined for both known and potential new
sources. Consistency checks are necessary to eliminate spurious
artefacts. In addition to determining the detection significance, the
correct source shape (Gaussian in approximation) and the presence or
absence of systematic background structures in the maps were
verified. In order to accept a source detection in this work, a
significance above 3$\sigma$ in at least two (consecutive) energy
intervals was required.  Source detection significances ($\simeq F/dF$
for a background-dominated photon distribution) were determined in the
mosaics for known sources at their catalog positions, and for
candidate new sources at the positions in the sky maps with the
highest detection significances.

For each source and energy band the time-averaged count rate was
determined by averaging the count rates from each deconvolved ScW,
weighted by the variance.

For converting subsequently count rates into flux values, the source
spectra were normalized to the known \object{Crab} total spectrum
(nebula and pulsar), using a $5 \times 5$ rectangular dithering
observation of the Crab performed during revolution 102 as
reference. The source fluxes could, thus, be expressed in Crab units
and converted in photon fluxes using the average Crab power-law
spectrum ($F(E) = \alpha E^{-\Gamma}$) with photon index $\Gamma =
2.108$ and normalization $\alpha = 9.59$ \pcs at 1~keV
\citep{Willingale01_crab}. This relation is valid throughout the
IBIS-ISGRI waveband up to $\sim$180 keV, as was verified by
\citet{Kuiper06_axps}. This method for deriving IBIS-ISGRI source
spectra enables us to determine the spectra without detailed knowledge
of the instrument response, which contained significant uncertainties
at the time of our analyses.

Spectral-model fitting was performed with XSPEC version 11.3.1
\citep{Arnaud96_xspec}. For power-law fitting the PEGPWR model was
used, which is a normal power law, but the program returns the
power-law index and the unabsorbed power-law flux ($F_{{\rm
power\,law}}$) in a chosen energy band instead of a normalization at a
given energy to describe the spectrum.  This model has been chosen in
order to minimize the covariance between the parameters, resulting in
the best error estimates.

All errors quoted in this paper are $1\sigma$ errors, unless stated
otherwise.

\section{Sky maps and source list}
\begin{table}
\renewcommand{\tabcolsep}{1.0mm}

\caption[]{List of all detected sources in the Cassiopeia region
ordered by right ascension.  Most sources are High-Mass X-ray Binaries
(HMXBs), either a Be X-ray Binary (BeXB) or a Supergiant X-ray Binary
(SXB), with in addition one Intermediate Polar (IP), one BL-Lac type
AGN, one Anomalous X-ray Pulsar (AXP) and two supernova remnants
(SNR). }

\begin{tabular}{l r l@{ $\pm$ }l l@{ $\pm$ }l l}
\vspace{-3mm}\\

\hline
\hline

Source & Exp. & \multicolumn{2}{l}{Flux (mCrab)}& 
\multicolumn{2}{l}{Flux (mCrab)}& Class\\
       & (ks)     & \multicolumn{2}{l}{20--50 keV}  & 
\multicolumn{2}{l}{50-100 keV} &\\

\hline

IGR~J00234+6141       & 1606 &  0.72&0.12 & 1.4&0.3       & HMXB?\\
Tycho                 & 1606 &  0.43&0.12 & 1.1&0.3$^{b}$ & SNR\\
V709~Cas              & 1606 &  3.84&0.12 & 2.8&0.4       & IP\\
1ES~0033+595          & 1606 &  0.69&0.13 & 1.2&0.4$^{b}$ & BL-Lac\\
IGR~J00370+6122       & 1606 &  0.78&0.13 & 1.1&0.4$^{b}$ & SXB\\
IGR~J00370+6122$^{a}$ &  489 &  2.7 &0.2  & 3.6&0.7            \\
Gamma~Cas             & 1602 &  3.8 &0.2  & 1.2&0.4$^{b}$ & BeXB\\
2S~0114+650           & 1579 &  7.8 &0.2  & 4.1&0.5       & SXB\\
2S~0114+650$^{a}$     &  936 & 12.1 &0.2  & 6.1&0.6            \\
4U~0142+61            & 1152 &  1.4 &0.3  & 7.0&0.7       & AXP\\
RX~J0146.9+6121       & 1133 &  1.6 &0.3  & 1.5&0.8$^{b}$ & BeXB\\
4U~2206+54            &  731 &  2.8 &0.5  & 1.8&1.2$^{b}$ & BeXB\\
Cassiopeia~A          & 1609 &  3.54&0.13 & 1.8&0.4       & SNR\\

\hline
\end{tabular}

\label{tab:src}
1 mCrab (20--50 keV)$= 6.66 \times 10^{-6}$ \pcsk,
1 mCrab (50--100 keV)$= 1.22\times 10^{-6}$ \pcsk;
$^{a}$ optimized data selection; $^{b}$ possibly contaminated by
artefact in map. Values to be used with care or used as upper 
limits.
\end{table}

Figures~\ref{fig:field20-50} and \ref{fig:field50-100} show sky maps
of a large part of the Cassiopeia region in the energy bands 20--50
keV and 50--100 keV, respectively. For presentation purposes the sky
regions at higher Galactic latitudes and below Galactic longitude $l
\sim 108^{\circ}$ were cropped off. As a result 4U~2206+54, located at
$(l,b) = (100\fdg6,-1\fdg1)$, is not shown in
Figures~\ref{fig:field20-50} and \ref{fig:field50-100}. The
significantly detected sources that passed our consistency checks as
mentioned in Sect.~\ref{sec:obs} are indicated by white circles and
annotation. The yellow circles and annotation indicate weakly detected
sources, transient sources, that were detected by INTEGRAL at other
epochs and the second AXP in the field, 1E~2259+586, for which there
is so far no INTEGRAL detection. The weak (one energy interval)
detections and upper limits of the latter sources are given and
discussed in Sects.~\ref{sec:2259} and \ref{sec:nodet}. All (weakly)
detected sources are listed in Table~\ref{tab:src}.

\section{Anomalous X-ray pulsars}
\label{sec:axp}

Anomalous X-ray pulsars, originally discussed as a subset of low-mass
X-ray binaries \citep[e.g.][]{Hellier94_0142, Mereghetti95_axps}, are
members of a special subset of isolated neutron stars
\citep[see][\,for a recent review]{Woods05_review}. So far, there are
seven established AXPs and one AXP candidate.  Their spin periods
($P$) are found in a narrow range between 5.5 s and 11.8 s and their
period derivatives ($\dot P$) are of the order of $\sim$$10^{-11}$ s
s$^{-1}$. Their characteristic ages ($\tau \sim P/2\dot P$) indicate
that they are young objects, though the characteristic age should be
interpreted with care, especially for young neutron stars. However,
the facts that all the Galactic AXPs (one is located in the SMC) are
found within $\sim$$1^{\circ}$ of the Galactic plane and that three of
them are associated with supernova remnants \citep{Gaensler01_snr}
support their young nature. These pulsars are called anomalous,
because their X-ray luminosities below 10 keV are orders of magnitude
higher than the rotational energy losses. The magnetic field strengths
of AXPs, $B \propto \sqrt{P\dot P}$, are of the order $10^{14}$ G,
much stronger than estimated for normal (radio) pulsars ($B \sim
10^{12}$ G).

In order to explain the energy budget of the AXPs, the magnetar model
\citep{TD93,TD96} is currently the most accepted model. This model was
originally aimed at explaining phenomena observed in Soft Gamma-ray
Repeaters (SGRs), but due to recent observations of similar behaviour
of SGRs and AXPs, like bursting activity \citep{Gavriil02_1048,
Kaspi03_2259}, they are now believed to originate from the same class
of objects.  This model also explains the glitching phenomena detected
in the spin down of two AXPs \citep{Kaspi03_1708, DallOsso03_1708,
Woods04_2259}. The magnetar model postulates that AXPs and SGRs are
powered by magnetic-field decay.

In the pre-INTEGRAL era AXPs were known as soft X-ray sources visible
up to 10 keV. All AXP spectra, below 10 keV, could be described by a
black-body component and a soft power law. However, this view has
changed completely since INTEGRAL discovered hard X-rays above 20 keV
from the direction of AXPs 1E~1841-045 and 1RXS~J170849-400910
\citep{Molkov04_sagarm, Revnivtsev04_gc}. In addition,
\citet{denHartog04_atel0142} reported the INTEGRAL detection of
4U~0142+61, in a preliminary analysis of the data used in this
work. \citet{Kuiper04_1841} provided definite evidence that the hard
X-rays from the direction of 1E~1841-045 indeed originated from the
AXP rather than the associated supernova remnant G27.4+0.0 (Kes
73). They showed that a significant fraction of the hard X-ray
emission is pulsed, using RXTE PCA and HEXTE covering the same energy
range as INTEGRAL.  Also for AXPs 1RXS~J170849-400910 and 4U~0142+61
\citet{Kuiper06_axps} subsequently found pulsed emission in the hard
X-ray regime above 20 keV using archival RXTE data.

\subsection{4U~0142+61}
\label{ch:0142}

\begin{figure}
\psfig{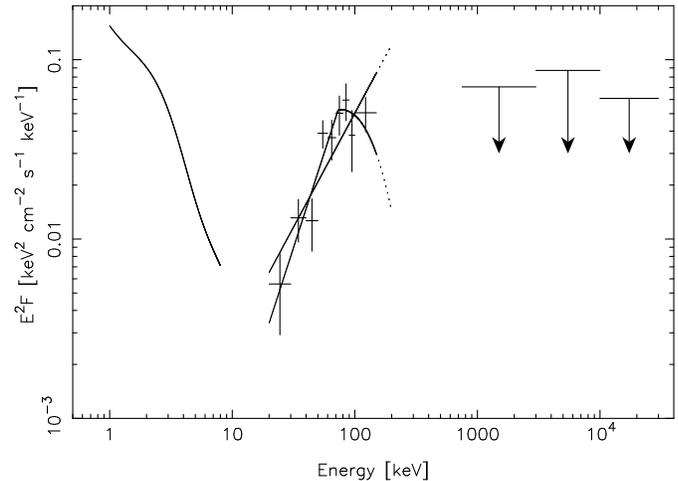}
\caption{The broad-band X-ray to gamma-ray spectrum of 4U~0142+61 in
 an $E^2 F$ representation. Shown are the 1--8 keV \emph{Chandra} fit
 \citep{Patel03_0142}, the 20--150 keV IBIS-ISGRI flux values
 (1$\sigma$ error bars) with 2 fits (see Table \ref{tab:0142}) and the
 0.75--30 MeV COMPTEL 2$\sigma$ upper limits. The dotted lines are
 extrapolations of the models fitted to the IBIS-ISGRI flux
 values. The extrapolation of the single power law shows that the
 spectrum has to bend/break in order not to be in conflict with the
 COMPTEL upper limits.
\label{fig:0142sp}}
\end{figure}

4U~0142+61 was detected by the {\em Uhuru} satellite in the early
seventies \citep{Giacconi72_uhuru, Forman78_uhuru}. The nature of
4U~0142+61 was unclear for more than two decades, caused by source
confusion with the nearby located high-mass X-ray binary
RX~J0146.9+6121 (24\arcmin \,from the AXP, see
section~\ref{sec:rx0146}).  With the discovery of RX~J0146.9+6121 in
ROSAT 0.1--2.4 keV data \citep{Motch91_0146},
\citet{Mereghetti93_0142} realized that the 25-minute periodicity, up
to then attributed to 4U~0142+61 \citep{White87_0142}, was unlikely to
originate from 4U~0142+61, because of the characteristics of the
modulations.  \citet{Hellier94_0142} showed with ROSAT PSPC
observations that this periodicity indeed originated from
RX~J0146.9+6121. The spin period of 8.7 seconds of 4U~0142+61 was
discovered by \citet{Israel94_0142}. They found that the rotational
energy loss of 4U~0142+61 is not sufficient to explain its X-ray
luminosity below 10 keV.

4U~0142+61 was the first AXP for which optical emission was discovered
\citep{Hulleman00_0142}, which later turned out to be pulsed
\citep{Kern02_0142}. \citet{Hulleman04_0142} also discovered near-IR
emission from this source, further studied by \citet{Morii05_0142} in
simultaneous X-ray, near-IR and optical observations. They discovered
in \emph{Subaru} observations, that also the near-IR emission is pulsed
\citep[][ private communication]{Morii06_0142}.  So far, no radio
emission has been detected \citep{Gaensler01_snr}.

The soft X-ray spectrum observed with \emph{Chandra} by
\citet{Patel03_0142} shows a typical AXP spectrum.  The spectrum can
be fitted with a thermal black-body model with a temperature $kT =
0.470 \pm 0.008$ keV and a very soft power-law component with a photon
index $\Gamma = 3.40 \pm 0.06$.

In this work no INTEGRAL-timing analysis has been performed for
4U~0142+61.  This will be addressed in future work, exploiting
also additional INTEGRAL observations.

\subsubsection{4U~0142+61: INTEGRAL results}
\label{ch:0142res}

4U~0142+61 was significantly detected by INTEGRAL as a hard X-ray
source \citep[see Figs.~\ref{fig:field20-50} and \ref{fig:field50-100};
for a preliminary report see][]{denHartog04_atel0142}, even though
this source is located $17\fdg7$ from Cassiopeia~A and $9\fdg5$ from
Tycho, therefore being always outside the fully-coded FOV of
IBIS-ISGRI. In the sky-image mosaic with a total exposure of 1.15 Ms,
4U~0142+61 is detected with $5.0\sigma$, $9.5\sigma$ and $4.6\sigma$
significances in the energy bands 20--50 keV, 50--100 keV and 100--150
keV, respectively.

\begin{figure}
\psfig{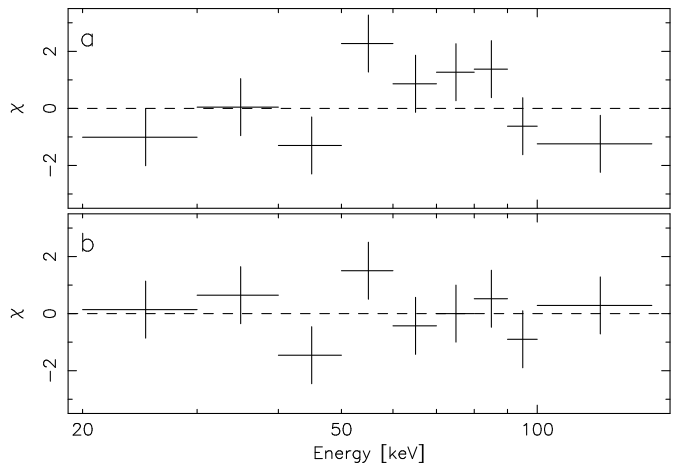}
\caption{ Fit residuals for the 4U~0142+61 spectrum: {\emph a}) for
a single power-law fit; {\emph b}) for a power-law with a
high-energy cutoff fit.
\label{fig:0142res}}
\end{figure}

The IBIS-ISGRI spectrum for energies above 20 keV, shown in
Fig.~\ref{fig:0142sp}, can be described by a power-law model with a
photon index $\Gamma = 0.73 \pm 0.17$. This fit has a \chir = 2.01
(dof = 7) and thus a null-hypothesis probability of 5\% (see
Table~\ref{tab:0142} and Fig.~\ref{fig:0142res}a for the fit
residuals). Striking is the remarkable hardening of the spectrum above
10 keV when compared to the very-soft \emph{Chandra} spectrum, also shown in
Fig.~\ref{fig:0142sp}.  The transition occurs between the \emph{Chandra} and
IBIS-ISGRI energy windows where it can be studied with RXTE
\citep{Kuiper06_axps}.

\subsubsection{4U~0142+61: COMPTEL archival-data analysis}
\label{ch:0142comptel}

The Compton telescope COMPTEL \citep{Schoenfelder93_comptel} aboard
the Compton Gamma-Ray Observatory (CGRO, 1991--2000) is sensitive to
gamma-ray photons between 0.75 MeV and 30 MeV, thereby covering the
harder gamma-ray band adjacent to the INTEGRAL one.  The very-hard
spectrum measured with IBIS-ISGRI for 4U~0142+61 prompted us to
reanalyze the COMPTEL data archive to search for a signal from 4U
0142+61. COMPTEL has a wide circular field of view covering $\sim$1
steradian. Therefore, during the long mission lifetime of CGRO, most
of the sky, particularly the Galactic plane, has been viewed with long
exposures. Sky maps and source parameters can be obtained with the
maximum-likelihood method \citep{deBoer92_comptel}, which is
implemented in the standard COMPTEL data-analysis package.  For the
analysis and data selections the approach described by
\citet{Zhang04_comptel} was followed.  The effective on-axis exposure
time on this AXP, accumulated over the total mission, amounts to
$\sim$4.2 Ms.  Standard energy intervals were selected for the
analysis. In the lower-energy intervals, 0.75--3 MeV and 3--10 MeV,
weak, low significance ($\la$$1.5\sigma$) excesses were found. In the
10--30 MeV energy band no positive-flux measurement was derived. The
$2\sigma$ upper limits for these energy intervals are $3.1 \times
10^{-8}$ \pcsk, $2.9 \times 10^{-9}$ \pcsk and $2.0 \times 10^{-10}$
\pcsk, respectively. These are shown in Fig.~3, and can be directly
compared with the hard X-ray spectrum. If the power-law spectrum
fitted to the IBIS-ISGRI data extends to the MeV energy range, it
would result in a photon flux of $\sim$$6.7 \times 10^{-7}$ \pcsk
(0.75--3 MeV), i.e. a bright COMPTEL source, inconsistent with the
COMPTEL upper limits. It shows that the spectrum has to bend or break
below $\sim$0.75 MeV. To search for possible indications of this break
in the IBIS-ISGRI data, a power-law model with a high-energy
cutoff was fitted to the data. This resulted in a fit improvement of
2.34$\sigma$ with a $\Delta \chi^2$ of 7.9 for 2 extra free parameters
(see Table \ref{tab:0142} and Fig.~\ref{fig:0142sp}).  This fit gives
an extremely hard power law in the lower energy part with a photon
index $\Gamma$ = {\bf --} 0.11 $\pm$ 0.43 and a high-energy cutoff at
73 $\pm$ 15 keV with an e-folding energy of 37 $\pm$ 18 keV. With this
model no significant flux is expected in the COMPTEL energy
range. Figure~\ref{fig:0142res}{\emph b} shows the residuals from the
fits. We merely consider this $2.3 \sigma$ fit improvement as a hint
for a break. With recently performed additional INTEGRAL observations
we expect to obtain tighter constraints.

\begin{table}

\caption[]{Spectral fit results for AXP 4U~0142+61. The data can be
fitted with a single power-law model. Using a power-law with a
high-energy cutoff model, the fit improves
2.34$\sigma$, indicating a hint for the expected break in the
spectrum.}

\begin{tabular}{l l}

\vspace{-3mm}\\

\hline
\hline

\multicolumn{2}{l}{Power law (20--150 keV)}\\
\hline
$\Gamma$ & $0.73 \pm 0.17$\\
$ F_{{\rm power\,law}}$ & $(9.7 \pm 0.9) \times 10^{-11}$ \ecs\\
\chir (dof) & 2.01 (7)\\
\hline

\multicolumn{2}{l}{Power law with High-Energy cutoff (20--150 keV)}\\
\hline
$\Gamma$               & $-0.11 \pm 0.43$\\
$F_{{\rm Power\,law}}$ & $(18 \pm 5) \times 10^{-11}$ \ecs\\
$E_{{\rm cutoff}}$     & $73 \pm 15$ keV\\
$E_{{\rm fold}}$       & $37 \pm 18$ keV\\
\chir (dof)            & 1.23 (5)\\
$F_{{\rm unabs}}$      & $9.5^{+0.5}_{-3.2} \times 10^{-11}$ \ecs\\
\hline

\end{tabular}

\label{tab:0142}
\end{table}

\subsection{1E~2259+586}
\label{sec:2259}

\begin{figure}
\psfig{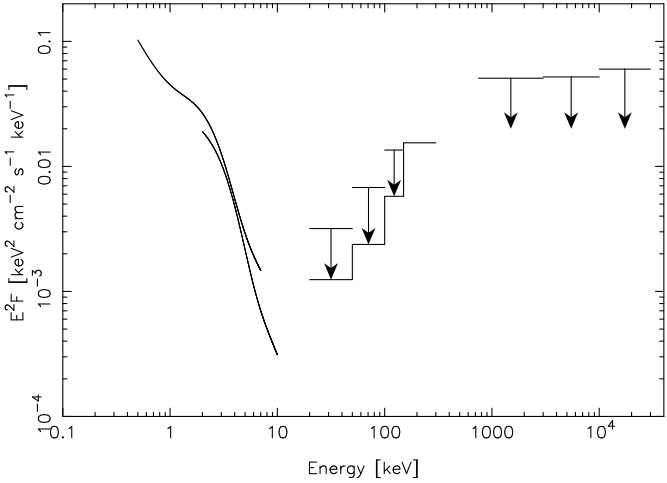}
\caption{Soft X-ray spectral fits and hard X-ray to gamma-ray upper
 limits of 1E~2259+586 in a $E^2 F$ representation. Shown are fits
 measured by \emph{Chandra} \citep[0.5--7 keV;][]{Patel01_2259} and by
 XMM-Newton \citep[2--10 keV; Obs. 1,][]{Woods04_2259}; the IBIS-ISGRI
 20--150 keV and COMPTEL 0.75--30 MeV $2\sigma$ upper limits. The
 solid staircase line is the IBIS-ISGRI $3\sigma$ sensitivity for 5-Ms
 exposure time. A hard high-energy spectral component as found for
 4U~0142+61 is not in conflict with the INTEGRAL upper limits.
\label{fig:2259sp}}
\end{figure}

1E~2259+586 has been very important for AXP research. Not only was it
the first of the currently known AXPs recognized as a peculiar source
\citep{Fahlman81_2259}, but it also played a key role in unifying AXPs
and SGRs as magnetar candidates when this source showed more than 80
SGR-like bursts in 2002 \citep{Kaspi03_2259}. The soft X-ray spectrum,
which can be described with a black body with a temperature $kT =
0.41$ keV and a very soft power-law with photon index $\Gamma = 3.6$,
is a typical AXP spectrum \citep[see][\,for recent X-ray
observations]{Patel01_2259}. Interestingly, \citet{Woods04_2259}
observed significant spectral changes (for energies below 10 keV)
before and after the 2002 outburst.

Contrary to 4U~0142+61, 1E~2259+586 is associated with a young
supernova remnant G109.1-1.0 (CTB 109), which was discovered by
\citet{Gregory80_2259} \citep[see also][]{Rho97_2259, Patel01_2259,
Sasaki04_2259}. The supernova remnant does not influence the INTEGRAL
analysis as it is only detected up to $\sim$4 keV
\citep{Sasaki04_2259}.

\subsubsection{1E~2259+586: INTEGRAL results}
1E~2259+586 is located only $2\fdg9$ from Cassiopeia~A (see
Fig.~\ref{fig:field20-50}). Therefore it was optimally exposed for 1.6
Ms in the fully-coded FOV. Even with the current high exposure, the
sensitivity of INTEGRAL is not sufficient to detect this AXP above 20
keV. Only upper limits could be extracted from the IBIS-ISGRI data. In
the energy bands 20--50 keV, 50--100 keV and 100--150 keV the $2
\sigma$ upper limits are $3.2 \times 10^{-6}$ \pcsk, $1.4 \times
10^{-6}$ \pcsk and $0.9 \times 10^{-6}$ \pcsk, respectively. Given the
reported variability below 10 keV, the data were also analyzed in
separate revolutions, but 1E~2259+586 was not detected.

\subsubsection{1E~2259+586: COMPTEL archival-data analysis}

Also for 1E~2259+586, analysis of COMPTEL archival data is performed
as described for 4U~0142+61 in Sect.~\ref{ch:0142comptel}.  The
effective on-axis exposure time on 1E~2259+586, accumulated over the
total mission, amounts to $\sim$4.9 Ms.  As for 4U~0142+61 COMPTEL
was not sensitive enough to significantly detect this AXP. The
$2\sigma$ upper limits for the 0.75--3 MeV, 3--10 MeV and 10--30 MeV
bands are $2.3 \times 10^{-8}$ \pcsk, $1.7 \times 10^{-9}$ \pcsk and
$2.0 \times 10^{-10}$ \pcsk, respectively.  The $2\sigma$ upper limits
are shown in Fig.~\ref{fig:2259sp}, together with the IBIS-ISGRI upper
limits and spectral fits to two different spectra measured below 10
keV by \emph{Chandra} \citep{Patel01_2259} and by XMM-Newton
\citep{Woods04_2259}. A spectral shape including a hard spectral
tail like measured for 4U~0142+61 (Fig.~\ref{fig:0142sp}), is allowed
by all upper limits above 10 keV.

Recently \citet{Kuiper06_axps} detected pulsed emission up to
$\sim$24~keV with the RXTE-PCA, indicating the likely onset of a hard
tail, which is still consistent with the upper limits derived in this
work. This detection is promising for a future INTEGRAL detection. The
total exposure on 1E~2259+586 will be more than 5 Ms by the end of
2006.

\section{Binaries}
\label{sec:binaries}
\begin{figure}
\psfig{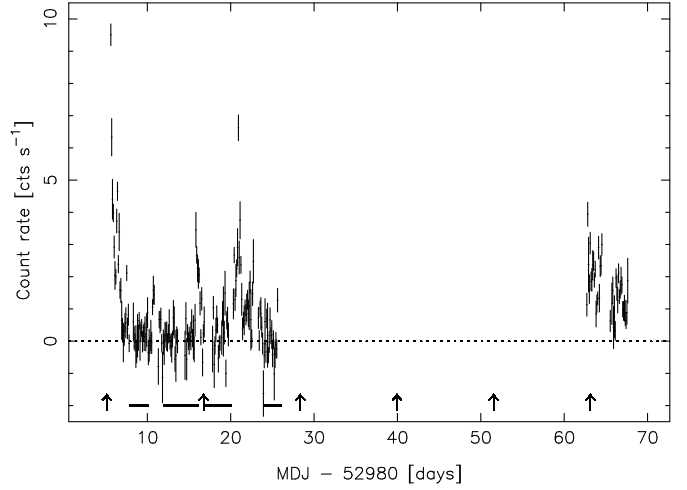}
\caption{IBIS-ISGRI (20--50 keV) light curve of 2S~0114+650 with 0.1
  day binning. It shows periods of flaring as well as quiescent
  phases.  The epochs of maximum flux, using an ephemeris based on
  RXTE-ASM data, are indicated with arrows. The four time intervals
  that are omitted for the spectral analysis are indicated as thick
  horizontal bars.
\label{fig:0114lc}}
\end{figure}

Since launch, INTEGRAL has discovered 107 new hard X-ray
sources\footnote{see
http://isdc.unige.ch/$\sim$rodrigue/html/igrsources.html for the
complete IGR-sources list}, among which a large number of binaries
(see e.g. \citet{Revnivtsev04_gc, Revnivtsev04_gccat, Molkov04_sagarm,
Molkov05_sagarmcat, Bird04_igrcat1, Bird06_igrcat2} and references
therein). One of the important discoveries of INTEGRAL is the
discovery of a previously missed class of objects, namely highly
absorbed High-Mass X-ray Binaries (HMXBs) \citep[$N_{\rm{H}} \gtrsim
10^{23} {\rm cm}^{-1}$; see][\,\,for a recent
review]{Kuulkers05_HMXB}.

HMXBs can be divided in two classes \citep{Corbet86}: the Be X-ray
binaries (BeXBs) and the supergiant X-ray (SXBs) binaries, which have
an OB supergiant or a bright giant as donor star. Currently there are
about 150 HMXBs known, out of which $\sim$65 are located in our own
Galaxy. Of these 150 systems, about 20 are of the supergiant class.
From the new INTEGRAL sources, for which follow-up observations are
performed, already 10 are identified as supergiant X-ray binaries with
a high intrinsic absorption. Including the new INTEGRAL SXBs, the
sample of SXBs has grown to about 30. See
e.g. \citet{Negueruela05_hmxb, Negueruela06_hmxb} and 
\citet{Lutovinov06_hmxb} for the latest developments and implications
of the newly found HMXBs by INTEGRAL.

In this section the binaries that were detected in the Cassiopeia
field (see Table~\ref{tab:src}) are introduced and the results are
presented. It concerns four HMXBs, one
Intermediate-Polar (IP) and the new source IGR~J00234+6141
which is most likely an HMXB.

\subsection{2S~0114+650}
\label{sec:0114}

2S~0114+650 is a supergiant X-ray binary. The supergiant companion
\object{LSI~ $+65^\circ 010$} is of the B1Ia type \citep{Reig96_0114}.  This
system exhibits X-ray modulations with a period of 2.7~hours
\citep{Finley92_0114,Corbet99_0114, Bonning05_0114} which are
interpreted as originating from the neutron star's spin - the slowest
known to date. The orbital period of 11.59~days
\citep{Crampton85_0114} is quite normal for a SXB. Recently,
\citet{Farrell04_0114} found evidence for a super-orbital period of
30.7~days, which they interpret as an indication for the presence of a
precessing warped accretion disk.

2S~0114+650 has been observed in the soft X-ray band several times
\citep[$<$10 keV; e.g.][]{Hall00_0114,Apparao91_0114}. However, before
INTEGRAL little information was available for the high-energy part of
the spectrum. \citet{Koeningsberger83_0114} published an OSO-8
spectrum taken in 1976, which could be fitted with a hard power law
with photon index $\Gamma = 1.2$ and a high-energy cutoff $E_{cut}
\sim$14 keV. The source was detected up to $\sim$30 keV.

\subsubsection{2S~0114+650: INTEGRAL results}
\begin{figure}
\psfig{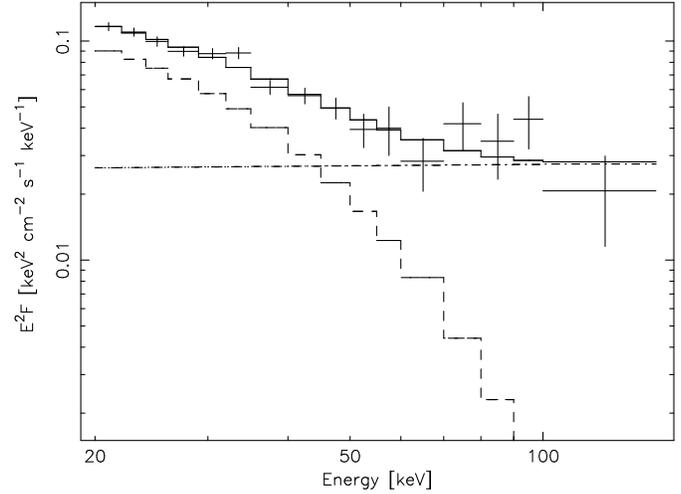}
\caption{Time averaged IBIS-ISGRI 20--150 keV spectrum of 2S~0114+650
  in an $E^2F$ representation.  This spectrum shows a high-energy tail
  which was not observed before INTEGRAL. The best fit to the data is
  a bremsstrahlung with a power-law model (see
  Table~\ref{tab:0114}). The best-fit model and the separate
  components are plotted.
\label{fig:0114sp}}
\end{figure}

INTEGRAL observed 2S~0114+650 for 1.58 Ms. It was detected at
significance levels of $43.1\sigma$ and $8.6\sigma$ in the 20--50 keV
and 50--100 keV bands, respectively. The light curve, presented in
Fig.~\ref{fig:0114lc}, shows a variable source with periods of flaring
and quiescence, both lasting of the order of less than a day to days.
This source is known to show flaring activity adding up to a single
broad pulse when folding with the orbital ephemeris over a long time
span \citep{Koeningsberger83_0114, Apparao91_0114,Corbet99_0114,
Hall00_0114}.  Using an ephemeris (maximum flux expected at MJD =
$51825\fd2 \pm 0\fd3 + N \times 11\fd5995(53)$) created with the
available online RXTE-ASM data (MJD 50087--53100), we find that the
phases of the flares are distributed consistently over the phase
interval of the broad time-averaged pulse.

\begin{figure}
\psfig{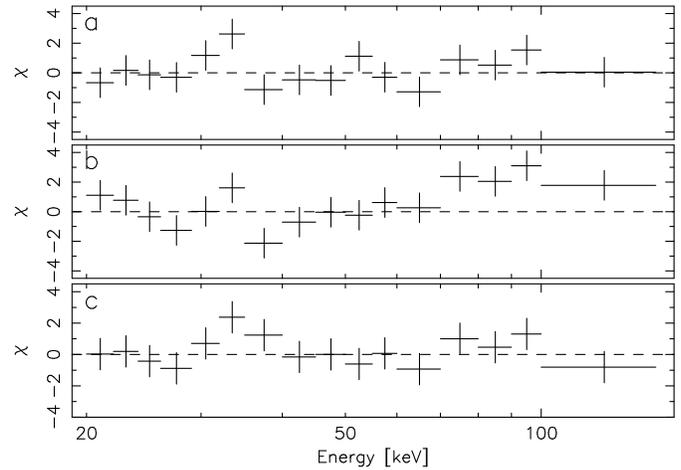}
\caption{Residuals for IBIS-ISGRI 20--150 keV spectral fits of
  2S~0114+650.  Panel {\emph a}) single power-law model; {\emph b})
  single bremsstrahlung model; {\emph c}) bremsstrahlung plus
  power-law model.
\label{fig:0114res}}
\end{figure}

For the spectral analysis, four time intervals in which the source was
in a weak/quiet state are omitted. The omitted intervals in MJD are:
52988--52990, 52992--52996, 52997--53000 and 53004--53006 (these
intervals are indicated in Fig.~\ref{fig:0114lc}). The remaining data
consist of 374 ScWs adding up to an exposure of 935.8 ks. For this
selection of observation times the detection significances increased
to $52.3\sigma$ and $10.0\sigma$ in the 20--50 keV and 50--100 keV
bands, respectively.

The time-averaged high-energy spectrum (20--150 keV) of 2S~0114+650 is
shown in Fig. \ref{fig:0114sp}. A fit with a single power-law model
($\Gamma = 3.01 \pm 0.06$) gives an acceptable fit up to 150 keV (see
Table \ref{tab:0114} and Fig.~\ref{fig:0114res}{\emph a}). A more
physical thermal-bremsstrahlung model could not fit the data
satisfactorily up to 150 keV. Above 70 keV it underestimates the
high-energy flux and the fit is therefore not acceptable (see
Table~\ref{tab:0114} and Fig.~\ref{fig:0114res}{\emph b}). Adding a
power-law component to the bremsstrahlung model, a good fit is
achieved. However, statistically it can not be favoured over, nor
distinguished from the single power-law model (see
Table~\ref{tab:0114} and Fig.~\ref{fig:0114res}\emph{a,c}). To explain
the hardest part of the spectrum comptonization models are
invoked. The models by \citet[][ compTT]{Titarchuk94_comptt} and
\citet[][ compPS]{Poutanen96_compps} are able to model the high-energy
part of the spectrum better than the bremsstrahlung model, but the
parameters are poorly constrained. Therefore the high-energy tail
which has been observed by INTEGRAL for the first time can be
described by either comptonization in a thermal or non-thermal
plasma. It is not clear whether it originates in a hot corona or in a
putative accretion disk. The accretion-disk scenario, however, is less
likely for SXB systems.

Recently \citet{Filippova05_igrbin} and \citet{Bonning05_0114} have
independently observed 2S~0114+65 with
INTEGRAL. \citet{Filippova05_igrbin} caught the source in a low state
and measured a simple power law ($\Gamma = 2.3 \pm 0.4$) which is
marginally consistent ($2\sigma$) with our
findings. \citet{Bonning05_0114} observed 2S~0114+6 for 180 ks in
December 2004. They find similar results compared to
\citet{Hall00_0114} combining JEM-X and IBIS-ISGRI data. Their
best-fit model is a power law with a high-energy cutoff: photon index
$\Gamma = 1.6 \pm 0.5$ and high-energy cutoff $E_{cut} = 9 \pm 11$ keV
with an e-folding energy $E_{fold} = 22.1^{+12.1}_{-6.0}$ keV (90\%
errors). Fitting our data with these parameters, leaving the
normalization free, gives an unacceptable fit. This model also fails
to fit the hard tail and therefore underestimates the flux above
70~keV. Already in their spectrum it can be seen that the model is
below their data points above 70 keV, but just because of the larger
statistical errors their model was still acceptable to describe their
spectrum. \citet{Bonning05_0114} note the indication for a possible
cyclotron absorption feature at 44 keV. Our data do not confirm this
indication, despite the much longer exposure time.

Even more recently, \citet{Masetti05_0114} published results from a
BeppoSAX observation of 63~ks with the narrow field instruments,
sensitive between 0.1 keV and 300 keV. The spectrum of 2S~0114+65 is
determined in the 1.5--100 keV band. Their findings agree with the
findings of \citet{Bonning05_0114}, except that they also do not find
evidence for a cyclotron absorption line. Compared to our spectrum
their fit results also fail to detect the hard tail.

\begin{table}
\caption[]{Spectral fit results for 2S~0114+650.}

\begin{tabular}{l l}

\vspace{-3mm}\\

\hline
\hline

Power law (20--150 keV)\\
\hline
$\Gamma$               & $3.01 \pm 0.06$\\
$F_{{\rm power\,law}}$ & $(17.2 \pm 0.5) \times 10^{-11}$ \ecs\\
\chir (dof)            & 1.21 (14)\\

\hline

Bremsstrahlung (20--150 keV)\\
\hline
$kT$                 & $20.4 \pm 0.9$ keV\\
norm                 & $(5.9 \pm 0.4) \times 10^{-2}$ \\
\chir (dof)          & 2.45 (14)\\
$F_{{\rm unabs}}$    & $(14.9 \pm 0.4) \times 10^{-11}$ \ecs\\
\hline

\multicolumn{2}{l}{Bremsstrahlung + Power law (20--150 keV)}\\
\hline
$kT$                      & $14.1 \pm 2.4$ keV\\
norm                      & $(7.2 \pm 1.5) \times 10^{-2}$\\
$\Gamma$                  & 2.0 $\pm$ 0.7\\
$F_{{\rm power\,law}}$    & $(8.7 \pm 3.7) \times 10^{-11}$ \ecs\\
\chir (dof)               & 1.12 (12)\\
$F_{{\rm unabs}}$         & $(17.7 \pm 1.0) \times 10^{-11}$ \ecs\\
\hline

\end{tabular}

\label{tab:0114}
\end{table}

\subsection{$\gamma$~Cas}
\label{sec:gamcas}

$\gamma$~Cas is a bright B0.5 IVe star and represents the prototypical
Be star \citep{Secchi1867_gamcas}. It was discovered in X-rays by
\citet{Jernigan76_gamcas} and \citet{Mason76_gamcas}. The X-ray
characteristics of $\gamma$~Cas made it a peculiar star because its
X-ray luminosity is several times higher than measured for other Be
stars, but $\sim$20 times weaker than the weakest BeXB
\citep{Murakami86_gamcas}. Through radial velocity analysis of
spectral lines, it was found to be a binary \citep{Harmanec00_gamcas}.

Determining whether the nature of the compact companion is a neutron
star or a white dwarf was a challenge for a long time.
\citet{Rappaport82_gamcas} argued that the X-ray luminosity of
$\sim$$10^{33}$ erg s$^{-1}$ combined with a wide orbit, and the
hardness of its spectrum, support a neutron-star companion. However,
most authors agree that a white dwarf companion is more likely: its
spectrum can be fitted with a thermal optically thin plasma with an
extremely high temperature of 10--12 keV; a strong Fe emission line at
6.8 keV with an equivalent width of 280 eV is present
\citep{Murakami86_gamcas, Owens99_gamcas}; and also rapid fluctuations
on timescales of seconds hinting towards a white-dwarf companion are
detected \citep{Kubo98_gamcas, Smith98_gamcas}.

\citet{Owens99_gamcas} observed $\gamma$~Cas with BeppoSAX. The
spectrum of $\gamma$~Cas could be extracted up to $\sim$30 keV with
the HPGSPC and $\sim$40 keV with the PDS. The best fit to their X-ray
spectrum from 0.2--40 keV is achieved with a MEKAL model \citep{mekal}
with an additional Carbon line (at 0.3 keV), or a two component MEKAL
model. The temperature of the hot plasma in both models is $12.4\pm
0.6$ keV.

\subsubsection{$\gamma$~Cas: INTEGRAL results}

\begin{figure}
\psfig{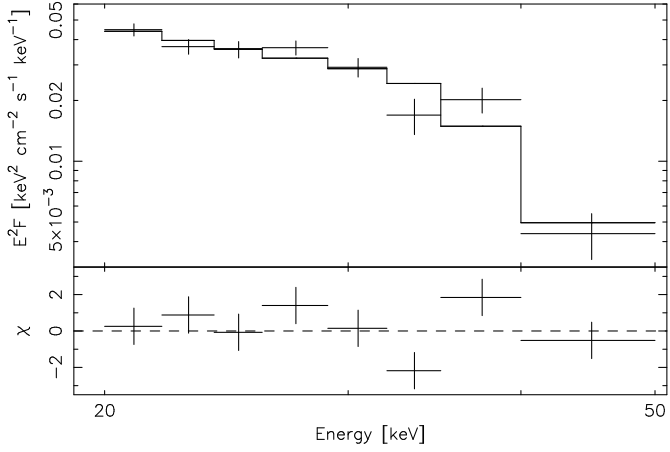}
\caption{IBIS-ISGRI spectrum of $\gamma$~Cas in an $E^2F$
representation. Overplotted is the best-fit model, which is a power-law
with a high-energy cutoff (see Table~\ref{tab:gamcas}). The bottom
panel shows the fit residuals.
\label{fig:gamcassp}}
\end{figure}

INTEGRAL detected $\gamma$~Cas with high significance up to 50 keV
($25.1\sigma$). Between 50 keV and 100 keV a 3$\sigma$ flux point is
measured (see Table~\ref{tab:src}), however, this flux point is likely
contaminated by surrounding artefacts. For this reason it was
discarded in the spectrum, shown in Fig.~\ref{fig:gamcassp}.

The INTEGRAL spectrum of $\gamma$~Cas can best be described by a
power-law model with a high-energy cutoff. However, the residuals are
large. With \chir = 2.81 (dof = 4) and thus a null-hypothesis
probability of 2.4\%, this fit can hardly be considered statistically
acceptable. However, other models, like a single power law (\chir =
7.32, dof = 6) or a bremsstrahlung model \mbox{(\chir = 4.77, dof = 7)},
result in significantly worse fits. This is due to the very low flux
in the 40--50 keV bin with respect to the 20--40 keV spectrum,
requiring a cutoff.

A bremsstrahlung fit over the 20--40 keV interval yields a temperature
of (13.6 $\pm$ 1.5) keV with \chir = 1.44 (dof = 5), which is (within
errors) comparable to the temperature found by
\citet{Owens99_gamcas}. However, over the 20--50 keV energy range the
best-fit temperature is $9.8 \pm 0.5$ keV. This temperature is in
agreement with reports by \citet{Kubo98_gamcas} and
\citet{Smith98_gamcas}, but the quality of the bremsstrahlung fit to
our spectrum is unacceptable.

The spectral properties of $\gamma$~Cas are more reminiscent of a
typical HMXB spectrum than a WD-binary spectrum, leaving the question
about the nature of the compact companion still open.

\begin{table}
\caption[]{Spectral fit results for $\gamma$~Cassiopeia.}

\begin{tabular}{l l}

\vspace{-3mm}\\

\hline
\hline

\multicolumn{2}{l}{Power law with High-Energy cutoff (20--50 keV)}\\
\hline
$\Gamma$               & $3.1 \pm 0.3$ \\
$F_{{\rm power\,law}}$ & $(4.2 \pm 0.3) \times 10^{-11}$ \ecs\\
$E_{{\rm cutoff}}$     & $34.1 \pm 1.9$ keV\\
$E_{{\rm fold}}$       & $7 \pm 2$ keV\\
\chir (dof)            & 2.81 (4)\\
$F_{{\rm unabs}}$      & $(3.56 \pm 0.14) \times 10^{-11}$ \ecs\\
\hline

Bremsstrahlung (20--40 keV)\\
\hline
$kT$                         & $13.6 \pm 1.5$ keV\\
norm                         & $3.6 \pm 0.7 \times 10^{-2}$ \\
\chir (dof)                  & 1.44 (5)\\
$F_{{\rm unabs}}$            & $3.4^{+20}_{-0.8} \times 10^{-11}$ \ecs\\
\hline

\end{tabular}

\label{tab:gamcas}
\end{table}

\subsection{RX~J0146.9+6121}
\label{sec:rx0146}

RX~J0146.9+6121 (\object{V* V831 Cas}), a BeXB located 24\arcmin
\,from AXP 4U~0142+61 (see section~\ref{ch:0142}), was discovered by
\citet{Motch91_0146} with ROSAT. Its companion star
(\object{LS~I~+61~235}) was spectrally classified as a B1Ve by
\citet{Reig97_0146}. \citet{White87_0142} discovered a 25 minutes
periodicity which was then attributed to AXP 4U~0142+61. However,
\citet{Hellier94_0142} proved that this periodicity originated from
RX~J0146.9+6121, using the ROSAT-PSPC imager. RX~J0146.9+612 also
shows long-term, up to an order of magnitude, variability
\citep{Haberl98_0146b}.

X-ray spectra derived from ASCA, RXTE and BeppoSAX data
\citep{Haberl98_0146, Haberl98_0146b, Mereghetti00_0146} could be
modelled up to $\sim$25 keV with power laws with or without high-energy
cutoffs or with bremsstrahlung models. However, the spectra derived
from the non-imaging instruments onboard RXTE and BeppoSAX are
contaminated by 4U~0142+61. For the soft X-ray spectrum ($<$10 keV) the
contribution of 4U~0142+61 was corrected for by including fixed model
parameters, but for the hard X-ray spectrum ($>$10 keV) the contribution of
4U~0142+61 was still unknown. The GIS instrument onboard ASCA was able
to resolve RX~J0146.9+6121 and 4U~0142+61 in the 0.5--10 keV
band.

\begin{figure*}[!ht]
\psfig{figure=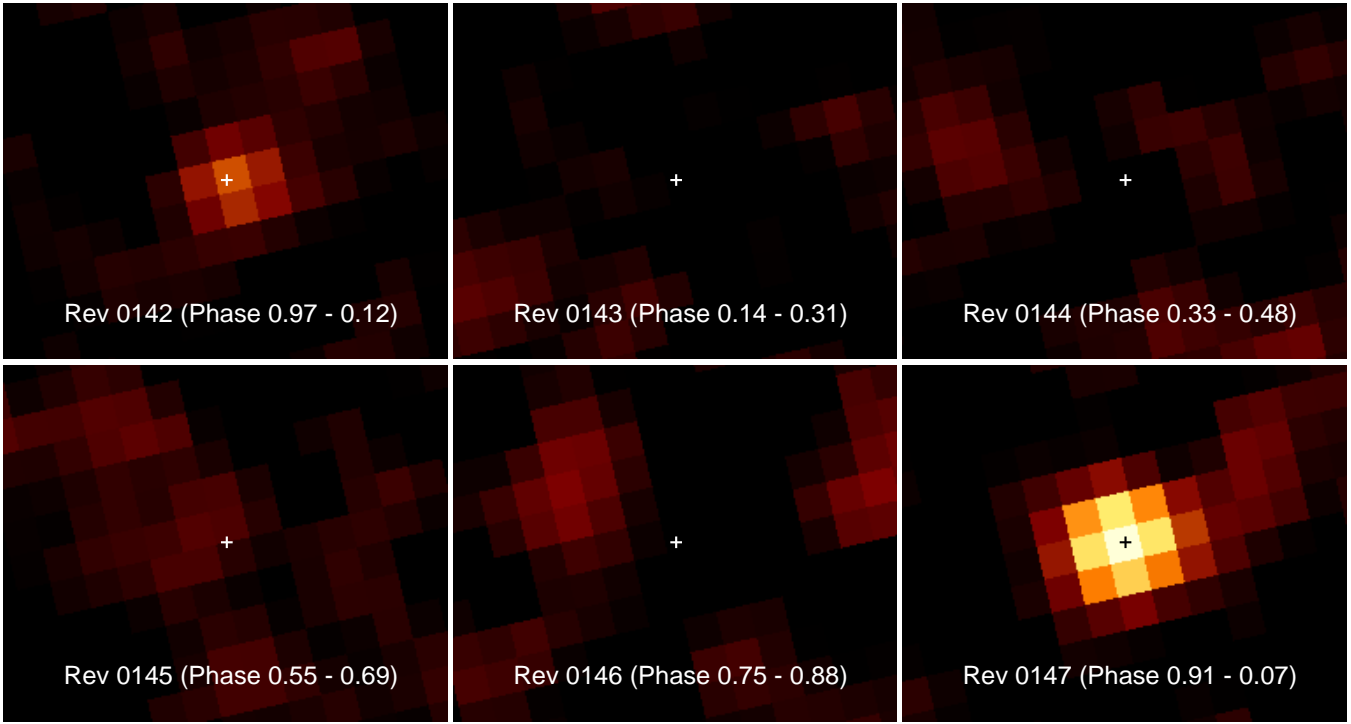,width=18cm,angle=-90,clip=t}
\caption{Series of 20--50 keV mosaics centered on the coordinates of
IGR~J00370+6122. The source was discovered in the revolution-147
mosaic as a $12.3\sigma$ source. In each panel the corresponding
orbital phases of each revolution are given. Both detections fall
within the expected epoch of maximum flux.
\label{fig:0037_panel}}
\end{figure*}

\subsubsection{RX~J0146.9+6121: INTEGRAL results}
Like AXP 4U~0142+61, this source was always in the partially-coded FOV
of IBIS-ISGRI with a total exposure of 1.13 Ms.  With the high spatial
resolution of INTEGRAL it is possible to resolve RX~J0146.9+6121 and
the AXP, and therefore there is no source confusion.  In the 20--50
keV mosaic the source is detected with a detection significance of
5.3$\sigma$ with a corresponding photon flux of $(1.1 \pm 0.2) \times
10^{-5}$ \pcsk. In narrow-band (5 keV) mosaics the source is only
detected in the 20--25 keV and 25--30 keV bands with photon fluxes
$(3.5 \pm 0.7) \times 10^{-5}$ \pcsk ($4.8\sigma$) and $(1.8 \pm 0.6)
\times 10^{-5}$ \pcsk ($3.3\sigma$), respectively. In the 30--40 keV
band RX~J0146.9+6121 is still (just) visible with $(0.8 \pm 0.3)
\times 10^{-5}$ \pcsk ($2.7\sigma$), but in the 40--50 keV band only a
2$\sigma$ upper limit of $4.2 \times 10^{-6}$ \pcsk could be
extracted.  We fitted a power-law model to these data points to obtain
an indication for the photon index and flux. The photon index is
$\Gamma = 3.4 \pm 0.9$ with a flux $(1.5 \pm 0.2) \times 10^{-11}$
\ecs (20--40 keV).

Comparing these results with the previously published results from
\citet{Haberl98_0146, Haberl98_0146b} and \citet{Mereghetti00_0146},
our data points nicely extend the previously observed spectra. Leaving
the normalization as the only free parameter, acceptable fits can be
found for all reported fits, except for one, namely the single
power-law fit with photon index $\Gamma = 1.46$ observed with ASCA
\citep{Haberl98_0146}, does not give an acceptable fit to our data
points. However, the bremsstrahlung model fitted to the same ASCA data
returns the best slope for the INTEGRAL energy band between 20 and 40
keV. The fluxes in the 0.5--10 keV band are also comparable with
$\sim$$3.9 \times 10^{-11}$ \ecs for ASCA and $(2.9 \pm 0.6) \times
10^{-11}$ \ecs for INTEGRAL (extrapolated towards lower energies). The
power-law with a high-energy cutoff model (\mbox{$\Gamma = 2.05$},
\mbox{$E_{{\rm cutoff}} = 17.2$}, $E_{{\rm fold}} = 7$), fitted to
RXTE data \citet{Mereghetti00_0146}, gives a slightly too steep
spectral slope in the INTEGRAL energy range. However, their
normalization of this power law is in full agreement with the towards
lower energies extrapolated INTEGRAL normalization of $(1.4 \pm
0.2)\times 10^{-10}$ \ecs.

Recently, \citet{Filippova05_igrbin} reported the detection of
RX~J0146.9+6121 with independent INTEGRAL observations. They reported
a weak detection of $\sim$3 mCrab (18--60 keV) and a spectrum fitted
with a single power law with photon index $\Gamma =
2.9^{+1.1}_{-0.8}$. The slopes of our spectra are fully consistent,
however during their observation the source was about twice as bright.

\subsection{IGR~J00370+6122}
\label{sec:0037}
\subsubsection{IGR~J00370+6122: Discovery and identification}

IGR~J00370+6122 is the first newly found INTEGRAL source in the
Cassiopeia region \citep{denHartog04_atel0037}. The source did not
show up in the integrated mosaics of revolutions 142--148 (1.6 Ms),
but it did appear in two single revolutions. For this paper the data
were reanalyzed with OSA 4.2, confirming our initial results (see
Fig.~\ref{fig:0037_panel}).  The source is significantly detected in
revolution 147 with a significance of $12.3\sigma$ and a flux of
$(3.1 \pm 0.3) \times 10^{-5}$ \pcsk (4.6 mCrab) in the 20--50 keV
energy band.  In the other revolutions the source is not detected
except in revolution 142 where it appeared more weakly with a
significance of $5.6\sigma$ and a flux of $(1.4 \pm 0.3) \times
10^{-5}$ \pcsk (2.1 mCrab, 20--50 keV).

We redetermined the source position in the 20--50 keV mosaic of
revolution 147: R.A. = $0^{\rm h} 37^{\rm m} 07\fs3$, Decl. =
$+61^{\circ} 21\arcmin 50\arcsec$ (J2000) with an estimated 2\arcmin\,
accuracy (90\% confidence), which is 0\farcm37 from the originally
distributed coordinates \citep{denHartog04_atel0037}.

A search through \emph{SIMBAD} yields a possible ROSAT X-ray
counterpart, 1RXS~J003709.6+612131 \citep{Voges99_rxs}, located only
0\farcm43 from the updated coordinates. The ROSAT source has an
optical counterpart within its 17\arcsec\, error radius, namely a
$\sim$10th-magnitude (B and V bands) B1Ib supergiant,
\object{BD~+60~73} \citep{Rutledge00_rosat}.  Follow-up optical
observations by \citet{Reig05_opthmxb} reassessed the spectral type of
BD~+60~73 to be BN0.5II-III, which belongs to the bright giant
class. The estimated distance to the source is $\sim$3 kpc. No radio
emission is found from this source in the NVSS online archive. Also
\citet{Cameron04_0037} did not find a radio signal with new VLA
observations.

\begin{figure}
\psfig{figure=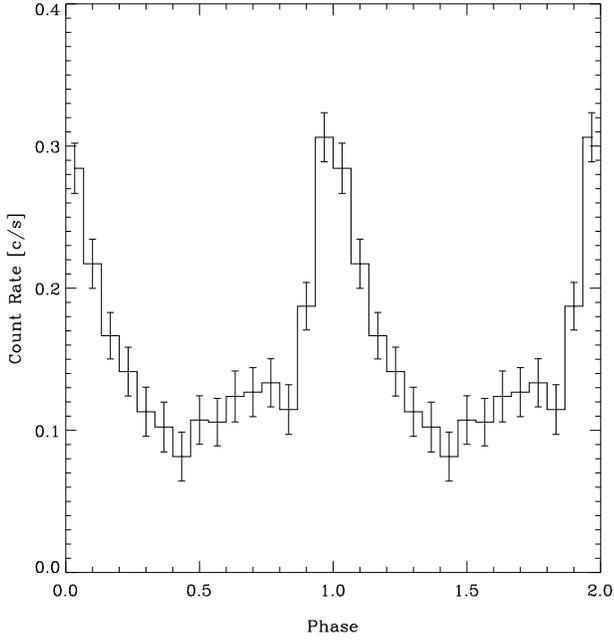,width=\columnwidth,angle=0,clip=t}
\caption{Folded RXTE-ASM light curve (1.1--12 keV) for
IGR~J00370+6122, including a $\simeq 0.07\, \rm{cts\, s^{-1}}$
systematic bias, showing a quiescent level below 1 mCrab and a peak
flux of 3.3 mCrab ($\simeq$$0.23\, \rm{cts\, s^{-1}}$). The source is
above its $75\%$ peak flux only for $15\%$ of the time, a time window
of $\sim$2.5 days per 15.665-day cycle. Two periods are
shown for clarity.
\label{fig:asm_orb}}
\end{figure}

We obtained a RXTE All-Sky Monitor (RXTE-ASM) light curve (MJD
 50087--53124) and found a periodicity of (15.665 $\pm$ 0.006) days.
The folded light curve in Fig.~\ref{fig:asm_orb} shows a single peak
with a peak flux of 3.3 mCrab and a quiescent level below 1 mCrab
(after subtracting a 1 mCrab systematic bias). The peak flux remains
above 2 mCrab for $\sim$3 days. The epoch of maximum flux during the
INTEGRAL and contemporaneous RXTE-ASM observations is \mbox{MJD
53001.7 $\pm$ 0.3}.  A HMXB with an eccentric orbit where the compact
object is accreting from wind accretion near periastron, can exhibit a
light curve like in Fig.~\ref{fig:asm_orb}. Therefore this periodicity
can be interpreted as the orbital period of a binary system.

Using this ephemeris we found that the detections during revolutions
142 and 147 fall within orbital phases 0.97--1.12 and 0.91--1.07,
respectively, both at epochs of expected maximum flux. This shows that
IGR~J00370+6122 is indeed the hard X-ray counterpart of the variable
source which was found with RXTE-ASM and is most likely
\object{1RXS~J003709.6+612131}. 

To derive an IBIS-ISGRI spectrum we selected only ScWs falling within
the orbital-phase window 0.8--1.2, resulting in 489 ks of
exposure. IGR~J00370+6122 is detected up to 80 keV. The spectrum is
shown in Fig.~\ref{fig:igr0037sp} and can be fitted with a single
power-law model with a photon index of $\Gamma = 1.9 \pm 0.3$ (see
Table~\ref{tab:0037}).

\subsubsection{IGR~J00370+6122: BeppoSAX archival observations}

BeppoSAX \citep{Boella97_bepposax}, operational for six years between
April 1996 and April 2002, observed this field with the Wide-Field
Cameras \citep[WFCs;][]{Jager97_wfc} serendipitously on a regular
basis, resulting in a total exposure of 1.37 Ms. We revisited these
data and detected IGR~J00370+6122 on four occasions in one-day
averaged time bins.  The spectral data were extracted in 31 channels
between 2 and 28 keV. Channels 1, 2 and 31 are not well calibrated and
were excluded. The 2--28 keV fluxes of the detections are $3.0 \pm
0.7$ mCrab (MJD 50840.1776--50841.3167), $8.7 \pm 1.4$ mCrab (MJD
50980.1576--50981.2459), $14.9 \pm 2.1$ mCrab (51135.8108--51136.4929)
and $4.6 \pm 0.5$ mCrab (MJD 51371.9518--51373.0377). All four
detections are within 0.1 orbital period from the expected peak as
derived from the ASM light curve. The WFCs have covered 23 more epochs
between orbital phase 0.9 and 1.1, but the source was not detected in
any of these observations with upper limits less than 2 mCrab. This
indicates the variable nature of the peak maxima. The spectrum of the
brightest maximum is shown in Fig.~\ref{fig:igr0037sp}. The data are
fitted with an absorbed power law ($\Gamma = 2.6 \pm 0.5$) (see
Table~\ref{tab:0037}).

\subsubsection{IGR~J00370+6122: RXTE ToO observations}

\begin{figure}[t]
\psfig{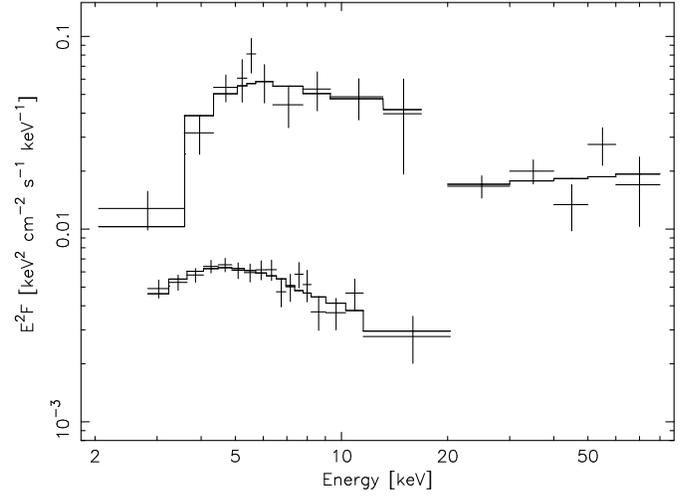}
\caption{Spectra (including absorption) in an $E^2F$ representation of
IGR~J00370+6122 measured with three instruments at three different
times: BeppoSAX Wide-Field Camera (2--17 keV, upper spectrum); RXTE
PCA (3--20 keV, lower spectrum); and INTEGRAL IBIS-ISGRI (20--80
keV). For the time-averaged IBIS-ISGRI spectrum only data within the
orbital phase bin 0.8--1.2 was used. The RXTE and BeppoSAX spectra
were taken around phases 0.91 and 0.02, respectively. Overplotted are
the best fits to the data which are described in
Table~\ref{tab:0037}. The order-of-magnitude flux difference between
the RXTE and BeppoSAX observations clearly shows the variable nature
of the peak luminosity.
\label{fig:igr0037sp}}
\end{figure}

In order to search for a possible X-ray pulsar we proposed a public
ToO observation with RXTE \citep{Bradt93_rxte}, which was
approved. Two short observations were performed on 2004-05-19 and
2004-06-02 (928 s and 3728 s; observation ID: 90415-01). The first
observation was performed near orbital phase 0.14, which is during the
decaying part of the peak. The second observation was performed near
orbital phase 0.02, which is at the maximum of the peak. In both
observations, however, the source was found to be in a low state with
fluxes of 1.4 and 1.1 mCrab, respectively. In Fig.~\ref{fig:igr0037sp}
the background subtracted spectrum of the longest RXTE observation is
shown. The data are modeled with an absorbed power-law ($\Gamma = 2.74
\pm 0.18$, see Table~\ref{tab:0037}). Within the errors the spectra
of both observations are similar.

\begin{table}
\caption[]{Spectral fit results for IGR~J00370+6122}

\begin{tabular}{l l}

\vspace{-3mm}\\

\hline
\hline

\multicolumn{2}{l}{INTEGRAL: Power law (20--80 keV)}\\
\hline
$\Gamma$                                & $1.9 \pm 0.3$\\
$F_{{\rm power\,law}}$& $(4.0 \pm 0.4) \times 10^{-11}$ \ecs\\
\chir (dof)                             & 1.62 (3)\\
\hline

\multicolumn{2}{l}{BeppoSAX: Power law (2--20 keV)}\\
\hline
$N_{{\rm H}}$              & $(13 \pm 6) \times 10^{22} {\rm cm^{-2}}$ \\
$\Gamma$                                & $2.6 \pm 0.5$\\
$F_{{\rm power\,law}}$& $(30 \pm 7) \times 10^{-11}$ \ecs\\
\chir (dof)                             & 1.00 (21)\\
\hline

\multicolumn{2}{l}{RXTE: Power law (3--20 keV)} \\
\hline
$N_{{\rm H}}$              & $(6.9 \pm 1.8) \times 10^{22} {\rm cm^{-2}}$ \\
$\Gamma$                         & $2.9 \pm 0.2$\\
$F_{{\rm power\,law}}$& $(1.84 \pm 0.13) \times 10^{-11}$ \ecs\\
\chir (dof)                      & 0.45 (39)\\
\hline

\end{tabular}

\label{tab:0037}
\end{table}

\subsubsection{IGR~J00370+6122: Variability and absorption}

The INTEGRAL, BeppoSAX and RXTE spectra in Fig.~\ref{fig:igr0037sp}
clearly show that the activity of the source varies widely from orbit
to orbit. We compare the three observations in the \mbox{5--20 keV}
band, to minimize effects due to possible absorption
variations. During the (brightest) BeppoSAX observation, the flux was
$(10.7 \pm 1.7) \times 10^{-11}$ \ecs, while during the RXTE
observation, the flux was an order of magnitude less with $(9.2 \pm
0.6) \times 10^{-12}$ \ecs. The (towards lower energies extrapolated)
flux estimated from the INTEGRAL observations falls in between the
BeppoSAX and RXTE observations with a flux of $(3.4 \pm 1.1) \times
10^{-11}$ \ecs.

Comparing the BeppoSAX and the RXTE spectra, i.e. the most extreme
flux measurements, show similar power laws, $\Gamma = 2.9 \pm 0.2$ for
the RXTE data and $\Gamma = 2.6 \pm 0.5$ for the BeppoSAX data. The
fitted absorption ($N_{\rm H}$) to the BeppoSAX data of $(13 \pm 6)
\times 10^{22}\, \rm{cm^{-2}}$ is a factor of two higher than the fitted
absorption to the RXTE data of $(6.9 \pm 1.8) \times 10^{22}\,
\rm{cm^{-2}}$. However, due to the large error on the BeppoSAX value
it is impossible to verify whether the intrinsic absorption differs
from one maximum to another.

\subsection{V709~Cassiopeia (V709~Cas)}
\label{sec:v709}

V709~Cas was recognized as an Intermediate Polar (IP) by
\citet{Haberl95_v709}. IPs represent a subclass of Cataclysmic
Variables (CVs) which consist of a weakly ($B~<~10^6$~G) magnetized
White Dwarf (WD) showing asynchronous rotation and accretion from a
Roche-lobe filling low-mass companion star \citep[see][ for a
review]{Patterson94_IP}. \citet{Haberl95_v709} found the WD spin
period of 312.8 seconds. The orbital period was determined to be 5.34
hours by {\citet{Bonnet-Bidaud01_v709}. IPs are hard X-ray
sources. The hard X-ray photons are believed to originate from the
post-shock region above the magnetic poles. Their X-ray spectra can be
described by a thermal optically-thin bremsstrahlung model with plasma
temperatures of tens of keV. \citet{deMartino01_v709} observed
V709~Cas with BeppoSAX and RXTE. The time-averaged spectrum can be
modeled with a plasma of temperature $kT$ = 27~keV, plus a reflection
component from the white-dwarf surface. The BeppoSAX PDS could observe
V709~Cas up to $\sim$70~keV. The spin modulation was observed up to
$\sim$25~keV with a decreasing amplitude at higher energies, which is
typical for IPs.

\begin{figure}
\psfig{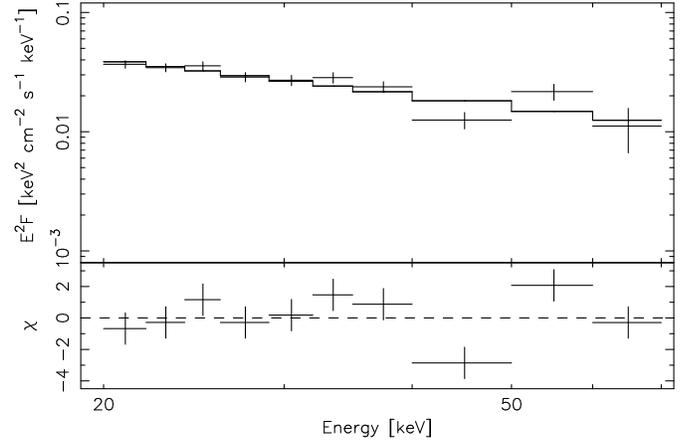}
\caption{IBIS-ISGRI spectrum of V709~Cas in an $E^2F$
representation. Over plotted is the best fit to the data, which is a
single power-law (see Table~\ref{tab:v709}). In the bottom panel the
residuals of the fit are shown.
\label{fig:v709sp}}
\end{figure}
\begin{table}
\caption[]{Fit results for V709~Cassiopeia}

\begin{tabular}{l l}

\vspace{-3mm}\\

\hline
\hline

Power law (20--70 keV) \\
\hline
$\Gamma$               & $3.00 \pm 0.12$\\
$F_{{\rm power\,law}}$ & $(4.62 \pm 0.18) \times 10^{-11}$ \ecs\\
\chir (dof)            & 2.18 (8)\\
\hline

Bremsstrahlung (20--70 keV)\\
\hline
$kT$              & $19.4 \pm 1.6$ keV\\
norm              & $(2.0 \pm 0.2) \times 10^{-2}$ \\
\chir (dof)       & 2.49 (8)\\
$F_{{\rm unabs}}$ & $(4.47 \pm 0.19) \times 10^{-11}$ \ecs\\
\hline

Bremsstrahlung (20--50 keV) \\
\hline
$kT$              & $17.1 \pm 1.5$ keV\\
norm              & $(2.4 \pm 0.3) \times 10^{-2}$ \\
\chir (dof)       & 1.37 (6)\\
$F_{{\rm unabs}}$ & $(3.82_{-0.28}^{+0.07}) \times 10^{-11}$ \ecs\\

\hline

\multicolumn{2}{l}{For all models the Galactic absorption was
fixed at}\\
\multicolumn{2}{l}{$0.46 \times 10^{22}\, {\rm cm^{-2}}$}

\end{tabular}

\label{tab:v709}
\end{table}

\subsubsection{V709~Cas: INTEGRAL results}
With a total exposure time of 1.6 Ms, V709~Cas is detected up to
$\sim$70 keV (see Fig.~\ref{fig:v709sp}). The best achieved fit of
this spectrum is a power-law model ($\Gamma = 3.00 \pm 0.12$, see
Table\ref{tab:v709}). However, with \chir = 2.18 (dof = 8) the fit is
not really acceptable. When fitted with a thermal-bremsstrahlung model
the fit is similarly unsatisfactory with \chir = 2.49 (dof = 8). The
high \chir is due to the flux values in the 40--50 keV and 50--60 keV
energy intervals, but there is no reason to discard these, since we
did not find similar discrepancies in our sample of source
spectra. Nevertheless, an acceptable fit can only be achieved when the
energy bins above 50 keV are ignored. The spectrum of V709~Cas in the
20--50 keV energy range can then be described by a
thermal-bremsstrahlung model (\chir = 1.37; \mbox{dof = 6}), with a
plasma temperature $kT = 17.1 \pm 1.5$ keV. However, the plasma
temperatures for the bremsstrahlung fits in both energy ranges are
comparable. Compared with the results of \citet{deMartino01_v709}, the
temperatures are significantly lower. However, they used a
bremsstrahlung model including a reflection component from the WD. A
reflection bump and a higher plasma temperature can mimic the spectral
slope of a simple bremsstrahlung model with a lower plasma
temperature. Fitting our spectrum with the spectral shape modeled by
\citet{deMartino01_v709} (model \emph{SAX \#2}; $kT = 27^{+6}_{-4}$
keV including a reflection component), leaving the normalization as
only free parameter, gives an unacceptable fit with \chir = 3.4 (dof =
9). The towards lower energies extrapolated unabsorbed flux (2--10
keV) of $(3.92 \pm 0.15) \times 10^{-11}$~\ecs is similar
($\sim$$12\%$ lower) to the flux measured with BeppoSAX. Adding the
plasma temperature as second free parameter, the fitted temperature is
$20.5 \pm 1.7$ keV (somewhat higher than without a reflection
component) with an unabsorbed flux $(5.4 \pm 0.7) \times 10^{-11}$
\ecs \mbox{(\chir = 2.8; dof = 8)}.

\citet{Falanga05_v709}, in a parallel study of a shorter IBIS-ISGRI
observation of V709~Cas (revolutions 262 and 263 for 181.9 ks),
derived a spectrum which they can also describe with a bremsstrahlung
model. Their plasma temperature ($kT = 25.5^{+9.3}_{-6.1}$ keV) is
higher, but statistically consistent with our results.

\subsection{IGR J00234+6141}
\label{sec:00234}
IGR~J00234+6141, the second new INTEGRAL source detected in the
Cassiopeia field, has no proper identification yet. The source was
detected in the mosaics for all available data, with a total exposure
of 1.6 Ms.  IGR~J00234+6141 is seen as a weak source ($0.72 \pm 0.12$
mCrab) with a significance of $5.9\sigma$ in the energy band 20--50
keV and with a significance of $4.1\sigma$ ($1.4 \pm 0.3$ mCrab) in
the 50--100 keV band, corresponding to $(4.8 \pm 0.8) \times 10^{-6}$
\pcsk and $(1.7 \pm 0.4) \times 10^{-6}$ \pcsk, respectively. The best
source position is R.A. = $0^{{\rm h}} 23^{{\rm m}} 24^{{\rm s}}$,
Decl. = +61\degr 41\arcmin 32\arcsec \,(J2000) with an estimated
3\arcmin \,accuracy (90\% confidence level), derived in the 20--30 keV
mosaic where the source is detected with $7.4\sigma$ significance.

Archival searches through {\em SIMBAD} yielded a possible ROSAT soft
X-ray counterpart, \object{1RXS~J002258.3+614111} \citep{Voges99_rxs},
which is located 3\farcm15 from the INTEGRAL centroid. It is
marginally consistent with the INTEGRAL position.  The RXTE-ASM
revealed a marginal detection, but no modulation was found in the
light curve. No reliable optical or IR counterpart can be found in the
INTEGRAL error box, because of the crowded field. Online available
radio data from NVSS showed no radio source.

Given its position close to the Galactic plane (b = -1\fdg0), the
lack of a radio counterpart and the nature of the majority of
previously reported new INTEGRAL sources near the Galactic plane, it
is more likely to be an X-ray binary, than an extragalactic source.

\section{Weakly and non-detected sources}
\label{sec:nodet}

In this section information can be found on additional sources that
have been weakly detected during this Cassiopeia-region survey, or
have been detected with INTEGRAL at other epochs. The sources
4U~2206+54 and 1ES~0033+595 are only detected in one energy band with
a significance greater than $5\sigma$. 4U~0115+63,
IGR~J00291+5434 and IGR~J01363+6610 were not detected during our
observations.

\begin{table}
\renewcommand{\tabcolsep}{1.7mm}
\caption[]{Fluxes and upper limits (UL) of weakly and non-detected sources. 
}

\begin{tabular}{l r@{ $\times$ }l l}

\vspace{-3mm}\\

\hline
\hline
\multicolumn{4}{c}{Source (Class): Exposure}\\
Energy band & \multicolumn{2}{l}{Flux (\pcsk)} & Significance\\

\hline
\multicolumn{4}{c}{4U~2206+54 (BeXB): 731 ks}\\

 20--30 keV & $(3.6 \pm 0.8)$&$10^{-5}$ & $4.7\sigma$\\
 30--40 keV & $(1.2 \pm 0.5)$&$10^{-5}$ & $2.4\sigma$\\
 
\hline

\multicolumn{4}{c}{1ES~0033+595 (BL-Lac): 1.6 Ms}\\

 20--30 keV & $(1.3 \pm 0.2)$&$10^{-5}$  & $6.3\sigma$ \\
 30--40 keV & $2.7$&$10^{-6}$  & $2\sigma$ UL\\

\hline

\multicolumn{4}{c}{4U~0115+63 (BeXB): 1.58 Ms}\\

 20--50 keV  & $2.7$&$10^{-6}$ & $2 \sigma$ UL\\
 50--100 keV & $1.2$&$10^{-6}$ & $2 \sigma$ UL\\
100--150 keV & $9.6$&$10^{-7}$ & $2 \sigma$ UL\\

\hline

\multicolumn{4}{c}{IGR~J00291+5434 (accreting msec pulsar):  1.6 Ms}\\

 20--50 keV   & $1.7$&$10^{-6}$ & $2 \sigma$ UL\\
 50--100 keV  & $4.3$&$10^{-7}$ & $2 \sigma$ UL\\
 100--150 keV & $7.2$&$10^{-7}$ & $2 \sigma$ UL\\

\hline

\multicolumn{4}{c}{IGR~J01363+6610 (BeXB): 1.1 Ms}\\

 20--50 keV  & $(5.6 \pm 1.5)$&$10^{-6}$ & $3.7\sigma^{(a)}$\\
 50--100 keV & $1.4$&$10^{-6}$ & $2 \sigma$ UL\\

\hline
$^{(a)}$ likely artefact\\
\end{tabular}

\label{tab:weak}
\end{table}

\subsection{4U~2206+54}
\label{sec:2206}

BeXB 4U~2206+54 is a hard X-ray source \citep{Giacconi72_uhuru} with
\object{BD~+53 2790} as optical counterpart
\citep{Negueruela01_2206}. The orbital period is 9.6 days
\citep{Corbet01_2206}. Observational evidence indicate that the
compact companion is most likely a neutron star \citep{Corbet01_2206,
Torrejon04_2206, Masetti04_2206, Blay05_2206}.

4U~2206+54 is located more than 11 degrees from
Cassiopeia~A. Therefore, the available exposure of 731 ks was
completely obtained in the partially-coded FOV of the IBIS-ISGRI
detector. Nevertheless, INTEGRAL has detected this source with a
significance of $5.7\sigma$ and a flux of $(1.8 \pm 0.3) \times
10^{-5}$ \pcsk (20--50 keV; see Tables \ref{tab:src} and
\ref{tab:weak}).  For a more detailed study of 4U~2206+54 with
INTEGRAL, we refer to \citet{Blay05_2206}.

\subsection{1ES~0033+595}
\label{sec:1es0033}

1ES~0033+595, an AGN of the BL-Lac type, was detected as a hard X-ray
source by BeppoSAX \citep{Costamante01_saxbllac}.  This source was
also found with INTEGRAL by \citet{Kuiper05_atelagn} with a photon
flux of $(1.1 \pm 0.2) \times 10^{-5}$ \pcsk (20--45 keV).  We
detected it in the 20--50 keV band, with $5.2\sigma$ significance (see
Tables \ref{tab:src} and \ref{tab:weak} for our flux values).
Comparing the high-energy fluxes of \citet{Costamante01_saxbllac} and
\citet{Kuiper05_atelagn} with our flux, shows the variable nature of
this BL Lac. While the source was a factor of 2.4 brighter during the
observations of \citet{Kuiper05_atelagn}, it was 4.7 times brighter
during the observations of \citet{Costamante01_saxbllac} with a photon
flux of $2.1 \times 10^{5}$ \pcsk (20--50 keV) in December 1999.

\subsection{Non detected 4U~0115+63, IGR~J00291+5434 and IGR~J01363+6610}
\label{sec:nondet}

In this field three other sources were detected by INTEGRAL at
other epochs.  4U~0115+63 was in outburst late August and September
2004 \citep{Lutovinov04_0115, Zurita04_0115}, reaching a peak flux of
about 400 mCrab. For INTEGRAL-ToO results we refer to
\citet{Ferrigno06_0115}. IGR~J00291+5434 was discovered by INTEGRAL
early December 2004 \citep{Eckert04_0029}. This new transient turned
out to be the fastest accretion-powered X-ray pulsar known to date
\citep{Markwardt04_0029,Markwardt04_0029_2}. More detailed INTEGRAL
and RXTE results are published by \citet{Shaw05_0029,Falanga05_0029}
and \citet{Galloway05_0029}.  IGR~J01363+6610 is a short transient
discovered with INTEGRAL by \citet{Grebenev04_01363} in April 2004.
The source was seen only once for 2.3 hours. We have not detected any
of these sources neither in the total exposure mosaics, nor in single
revolutions, nor in single SCWs. For flux upper limits see Table
\ref{tab:weak}.

\section{Summary}
\label{sec:disc}

We report results from a first deep observation of the Cassiopeia
region in the hard X-ray energy window of 20--200 keV, exploiting the
good imaging capabilities and the wide field of view of the IBIS-ISGRI
telescope aboard INTEGRAL.  We report results on nine 
detected sources of various kinds: Anomalous X-ray
Pulsar 4U~0142+61, several high-mass X-ray binaries, intermediate
polar V709~Cas, \mbox{BL-Lac} 1E~0033+595, and a still unidentified
new INTEGRAL source IGR J00234+6141. In addition the upper limits are
presented of the non-detected AXP 1E~2259+586 and three more high-mass
X-ray binaries.

The discovery in this work of the very hard spectral tail between 20
and 150 keV of AXP 4U~0142+61, increases the number of IBIS-ISGRI AXP
detections to three, the others being 1E~1841-045
\citep{Molkov04_sagarm, Kuiper04_1841} and 1RXS~J170849-400910
\citep{Revnivtsev04_gc}. This suggests that such a hard, highly
luminous spectral component is a common property of AXPs. Of these
three AXPs, 4U~0142+61 appears to exhibit the most extreme spectral
shape (see Fig.~\ref{fig:0142sp}): below 10 keV it has the softest
spectrum with a black-body temperature of $kT = 0.470 \pm 0.008$ keV
and a very soft power-law component with photon index $3.40 \pm
0.06$. Above 10 keV this drastically changes into a very hard
power-law component with photon index 0.73$\pm$0.17 (possibly even as
hard as $-0.11 \pm 0.43$ till $\sim$70 keV if the indication for a
break around that energy is real, see the fits in
Fig. \ref{fig:0142sp}). The X-ray luminosity between 20 and 100 keV
for the single power-law fit is $5.9 \times 10^{34}$ erg s$^{-1}$, a
factor of 440 higher than the rotational energy loss, amounting $1.3
\times 10^{32}$ erg s$^{-1}$ (assuming $M = 1.4\, \rm{M}_{\odot}$, $R
= 10$ km for a homogeneous sphere and a distance of \mbox{3 kpc}). Such a
high hard X-ray luminosity has to find its origin in the magnetosphere
of the magnetar, fed from the magnetic energy reservoir in the strong
magnetic field. The underlying scenario of production processes is not
yet well understood. To constrain different model scenarios, more
detailed observational parameters are needed. One of these is the
detailed spectral shape of the hard X-ray component. The reported
COMPTEL upper limits for energies between 0.75 MeV and 30 MeV already
indicate that the power-law spectrum of 4U 0142+61 has to bend/break
below this energy window. So far AXP 1E~2259+586 has not been detected
with INTEGRAL, however, the flux upper limits extracted from INTEGRAL
and COMPTEL data do not exclude the existence of a high-energy
tail. 

INTEGRAL has increased the number of known supergiant X-ray binaries
significantly \citep[see e.g.][]{Kuulkers05_HMXB, Negueruela06_hmxb,
Lutovinov06_hmxb}.  We introduce a new SXB, i.e. IGR~J00370+6122. We
detected this source in two INTEGRAL revolutions, at a position
consistent with that of the ROSAT source 1RXS J003709.6+612131, which
has a BN0.5II-III bright-giant counterpart. Analysis of archival
RXTE-ASM data revealed the $15.665 \pm 0.006$ day orbital
period. Follow-up studies using a ToO RXTE observation and
BeppoSAX-WFC archival data rendered a consistent picture of a new, by
an order of magnitude variable, wind-accreting SXB. The source
exhibits below 20 keV a power-law spectrum with index $\sim$2.7, and a
hard power-law spectrum with index $1.9 \pm 0.3$ in the INTEGRAL
window 20--80 keV.

We report the detection of a second SXB, 2S~0114+650. For the first
time the spectrum has been measured up to $\sim$150 keV. The good
statistics in the time-averaged spectrum above 70 keV, showed that a
single bremsstrahlung model cannot fit the data satisfactorily. In
addition, a power-law component with index $2.0 \pm 0.7$ is required,
or the spectrum can be explained with a single power law with index
$3.01 \pm 0.06$. 

Our deep survey of the Cassiopeia region rendered also detections of
three Be X-ray binaries: $\gamma$~Cas, RX~J0146.9+6121 and
4U~2206+54. For the first two, the INTEGRAL spectra were measured to
higher energies than published earlier, and are discussed in
comparison to these earlier reports. 

The intermediate polar V709~Cas was detected up to $\sim$70 keV.  The
best model fit was obtained for a power-law model yielding a photon
index of $3.00 \pm 0.12$, different from the bremsstrahlung-model fits
reported in literature.

Finally, two more hard-X-ray sources are found: 1) IGR~J00234+6141,
discovered as a weak ($\sim$1 mCrab) source up to 100 keV. No
counterpart could be found yet. Since it is located close to the
Galactic plane and there is no radio counterpart, it is likely an
X-ray binary, rather than an AGN. 2) 1ES~0033+595, an AGN of the
BL~Lac type, also detected in this work at the 1-mCrab level.

\begin{acknowledgements}
  We would like to thank R.H.D. Corbet, R. Remillard and the ASM/RXTE
  team for the rapid processing of the light curves for
  IGR~J00370+6122 and IGR~J00234+6141 and for useful discussions.  The
  results in this paper are based on observations with INTEGRAL, an
  ESA project with instruments and science data centre funded by ESA
  member states (especially the PI countries: Denmark, France,
  Germany, Italy, Switzerland, Spain), Czech Republic and Poland, and
  with the participation of Russia and the USA. We also acknowledge
  CDS SIMBAD and NRAO/VLA Sky Survey.

\end{acknowledgements}

\bibliography{literature}

\end{document}